\documentstyle[12pt]{article}
\def\be {\begin{equation}}
\def\ee {\end{equation}}
\def\ba {\begin{eqnarray}}
\def\ea {\end{eqnarray}}
\def\nn {\nonumber}
\def\r  {\rho}

\def\la {\label}
\def\le {\left}
\def\ri {\right}

\def\f {\frac}

\def\bi {\begin{itemize}}
\def\ei {\end{itemize}}
\begin{document}
\def\bea{\begin{eqnarray}}
\def\eea{\end{eqnarray}}
\title{\bf {Logarithmic correction to the Cardy-Verlinde formula in Achucarro-Oritz Black Hole }}
 \author{M.R. Setare  \footnote{E-mail: rezakord@ipm.ir}
  \\{Physics Dept. Inst. for Studies in Theo. Physics and
Mathematics(IPM)}\\
{P. O. Box 19395-5531, Tehran, IRAN }\\
 }

\maketitle
\begin{abstract}
In this paper we calculate leading order correction due to small
statistical fluctuations around equilibrium, to the
Bekenstein-Hawking entropy formula for the Achucarro-Oritz black
hole, which is the most general two-dimensional black hole derived
from the three-dimensional rotating Banados-Teitelboim-Zanelli
black hole. Then we obtain the same correction to the
Cardy-Verlinde entropy formula (which is supposed to be an entropy
formula of conformal field theory in any dimension).

 \end{abstract}
\newpage

 \section{Introduction}
 It is commonly believed that any valid theory of quantum gravity
 must necessary incorporate the Bekenestein-Hawking definition of
 black hole entropy \cite{bek,haw} into its conceptual framework.
 However, the microscopic origin of this entropy remains an enigma
 for two reasons. First of all although the various counting
 methods have pointed to the expected semi-classical result, there
 is still a lack of recognition as to what degrees of freedom are
 truly being counted. This ambiguity can be attributed to most of
 these methods being based on dualities with simpler theories,
 thus obscuring the physical interpretation from the perspective of
 the black hole in question. Secondly, the vast and varied number
 of successful counting techniques only serve to cloud up an
 already fuzzy picture.\\
 de Sitter/Conformal Field Theory correspondence (dS/CFT) \cite{Strom}-\cite{cad2} may hold
 the key to its microscopical interpretation. Naively, we would expect dS/CFT
 correspondence to proceed along the lines of Anti-de Sitter /Conformal Field Theory
  (AdS/CFT) correspondence \cite{mal} because de Sitter spacetime  can be
  obtained from anti-de Sitter spacetime by analytically continuing
  the cosmological constant to imaginary values. The Cardy-Verlinde formula proposed
   by Verlinde \cite{Verl}, relates the entropy of a  certain CFT with its total
energy and its Casimir energy in arbitrary dimensions. Using the
AdS$_{d}$/CFT$_{d-1}$ and dS$_{d}$/CFT$_{d-1}$ correspondences,
this formula has been shown to hold exactly for different black
holes. In previous paper \cite{set4}, by using the Cardy-Verlinde
formula, we have obtained the entropy of the Ach\'ucarro-Ortiz
black hole which is a two-dimensional black hole derived from the
three-dimensional rotating BTZ black hole.
\par In 1992
Ba\~nados, Teitelboim and Zanelli (BTZ) \cite{banados1,banados2}
showed that $(2+1)$-dimensional gravity has a black hole
 solution. This black hole is described by two (gravitational) parameters,
the mass $M$ and the angular momentum (spin) $J$. It is locally
AdS and thus it differs from Schwarzschild and Kerr solutions
since it is asymptotically anti-de-Sitter instead of flat
spacetime. Additionally, it has no curvature singularity at the
origin.
 AdS black holes, are members of this two-parametric family
of BTZ black holes and they are very  interesting in the framework
of string theory and black hole physics
\cite{strominger1,strominger2}.
\par For systems that
admit 2D CFTs as duals, the Cardy formula \cite{cardy} can be
applied directly. This formula gives the entropy of a CFT in terms
of the central charge $c$ and the eigenvalue of the Virasoro
operator $l_{0}$. However, it should be pointed out that this
evaluation is possible as soon as one has explicitly shown (e.g
using the AdS$_{d}$/CFT$_{d-1}$ correspondence) that the system
under consideration is in correspondence with a 2D CFT
\cite{CM99,andy}.\\
 In \cite{CM99} Cadoni and Mignemi, using Cardy
formula have been calculated the statistical entropy of
two-dimensional Jackiw-Teitelboim black hole, which can be
considered as the dimensional reduction of the $j=0$ (zero angular
momentum) BTZ black hole. Using a canonical realization of  the
asymptotic symmetry of two-dimensional anti-de Sitter space and
Cardy's formula they have been calculated the statistical entropy
of 2D black hole. In this case this reference relate a
two-dimensional black hole to a one-dimensional CFT, living on the
boundary of $AdS_2$. In fact the one-dimensional nature of the
boundary CFT, implies that we are dealing with some kind of
particle quantum mechanics, rather than quantum field theory. In
the other hand as have been shown in second paper by Cadoni and
Mignemi \cite{CM99} in the family of the AdS$_{d}$/CFT$_{d-1}$
dualities, the $d=2$ case is very similar to the $d=3$ one,  the
conformal group being in both instances infinite dimensional. But
a feature that is peculiar to the $d=2$ case is the complete
equivalence of the diffeomorphisms and the conformal group in  one
dimension. The physical implication of this equivalence is that
the usual difference between gauge symmetries and symmetries
related to conserved charges disappears. If one accepts the $d=2$
case is not fundamental then the CFT$_{1}$ should be thought of
just as (half) of CFT$_{2}$, in the way  have been explained in
\cite{CM99}, therefore in the $d=2$ context the general
AdS$_{d}$/CFT$_{d-1}$ duality becomes a duality between two 2d
conformal field theories. Although the possibility of describing
2D black holes by means of a CFT has been widely investigated
\cite{CM99,ads/cft,cadcav,carta1}, it is not completely clear if
it is always possible to mimic the gravitational dynamics of the
2D black hole through a CFT. However, in some cases and/or for
generic black holes in particular regimes, CFTs have been shown to
give a good description, this is in particular true for black
holes in AdS space. In this paper I consider the two-dimensional
Achucarro-Oritz black hole which is asymptotically AdS$_2$, then
we show a 2D CFT give a good description in this case also. When
the black hole is put in correspondence with a 2D CFT, the entropy
of black hole horizon is reproduce using the Cardy-Verlinde formula. \\
   There has been much recent interest in calculating the quantum
corrections to $S_{BH}$ (the Bekenestein-Hawking entropy)
\cite{maj1x}-\cite{set5}. The leading-order correction is
proportional to $\ln{S_{BH}}$. There are, {\it two} distinct and
separable sources for this logarithmic correction
 \cite{gg2x,maj3x} (see also recent paper by Gour and Medved \cite{ medved}).
  Firstly, there should be a correction
 to the number of microstates that is a quantum correction to the
 microcanonical entropy, secondly, as any black hole will typically exchange heat or
 matter with its surrounding, there should also be a correction due to thermal
 fluctuations in the horizon area. In a recent work Carlip
 \cite{carx} has deduced the leading order quantum correction to
 the classical Cardy formula. The Cardy formula follows from a
 saddle-point approximation of the partition function for a
 two-dimensional conformal field theory. This leads to the
 theory's density of states, which is related to the partition
 function by way of a Fourier transform \cite{carli}. In
 \cite{ajm1x}Medved has been applied the Carlip's formulation to
 the case of a generic model of two-dimensional gravity with
 coupling to a dilaton field.\\
 In this paper we consider the Achucarro-Oritz black hole, In section $2$ we calculate the
 corresponding thermodynamical quantities for  black hole horizon. In section $3$
  we calculate leading order correction due to small
statistical fluctuations around equilibrium, to the
Bekenstein-Hawking entropy formula then we obtain the same
correction to the Cardy-Verlinde entropy formula. In the other
term we assume the equality of Bekenstein-Hawking and CFT
entropy, then we uses the known statistical corrections to the
Bekenstein-Hawking entropy to predict the corrections to the
Cardy-Verlinde formula. Last section contain a summary of paper.

\section{Thermodynamical quantities of Achucarro-Oritz black hole}
The black hole solutions of Ba\~nados, Teitelboim and Zanelli
\cite{banados1,banados2} in $(2+1)$ spacetime dimensions are
derived from a three dimensional theory of gravity \be S=\int
dx^{3} \sqrt{-g}\,({}^{{\small(3)}} R+2\Lambda) \ee with a
negative cosmological constant ($\Lambda=\frac{1}{l^2}>0$).
\par\noindent
The corresponding line element is \be ds^2 =-\left(-M+
\frac{r^2}{l^2} +\frac{J^2}{4 r^2} \right)dt^2
+\frac{dr^2}{\left(-M+ \displaystyle{\frac{r^2}{l^2} +\frac{J^2}{4
r^2}} \right)} +r^2\left(d\theta -\frac{J}{2r^2}dt\right)^2
\label{metric}\ee There are many  ways to reduce the three
dimensional BTZ black hole solutions to the two dimensional
charged and uncharged dilatonic black holes \cite{ortiz,lowe}. The
Kaluza-Klein reduction of the $(2+1)$-dimensional metric
(\ref{metric}) yields a two-dimensional line element:
 \be ds^2 =-g(r)dt^2 +g(r)^{-1}dr^2
\label{metric1}\ee where \be g(r)=\left(-M+\frac{r^2}{l^2}
+\frac{J^2}{4 r^2}\right)\label{metric2}
 \ee
with $M$ the Arnowitt-Deser-Misner (ADM) mass, $J$ the angular
momentum (spin)
 of the BTZ black hole and $-\infty<t<+\infty$, $0\leq r<+\infty$, $0\leq \theta <2\pi$.
\par \noindent
The outer and inner horizons, i.e. $r_{+}$ (henceforth simply
black hole horizon) and $r_{-}$ respectively, concerning the
positive mass black hole spectrum with spin ($J\neq 0$) of the
line element (\ref{metric1}) are given as  \be
r^{2}_{\pm}=\frac{l^2}{2}\left(M\pm\sqrt{M^2 -
\displaystyle{\frac{J^2}{l^2}} }\right) \label{horizon1} \ee and
therefore, in terms of the inner and outer horizons, the black
hole mass and the angular momentum are given, respectively, by \be
M=\frac{r^{2}_{+}}{l^{2}}+\frac{J^{2}}{4r^{2}_{+}}\label{mass}\ee
and \be J=\frac{2\, r_{+}r_{-}}{l}\label{ang}\ee with the
corresponding angular velocity to be \be\Omega=\frac{J}{2
r_{+}^{2}}\label{angvel}\hspace{1ex}.\ee
\par\noindent The Hawking temperature $T_H$ of the black hole
horizon is \cite{kumar1} \bea T_H &=&\frac{1}{2\pi
r_{+}}\sqrt{\left(\displaystyle{\frac{
r_{+}^2}{l^2}+\frac{J^2}{4r_{+}^2}}\right)^2-\displaystyle{\frac{J^2}{l^2}}}\nn\\
&=&\frac{1}{2\pi r_{+}}\left(\displaystyle{\frac{
r_{+}^2}{l^2}-\frac{J^2}{4r_{+}^2}}\right)\label{temp1}
\hspace{1ex}.\eea \par\noindent In two spacetime dimensions we do
not have an area law for the black hole entropy, however one can
use thermodynamical reasoning to define the entropy \cite{kumar1}
\be S_{bh}=4 \pi r_{+} \label{entr1}\hspace{1ex}.\ee
 The specific heat of the black hole is given by
 \begin{equation}
 C=\frac{dE}{dT}=\frac{dM}{dT}=4\pi r_{+}(
 \frac{r_{+}^{2}-r_{-}^{2}}{r_{+}^{2}+3r_{-}^{2}})=S_{bh}(
 \frac{r_{+}^{2}-r_{-}^{2}}{r_{+}^{2}+3r_{-}^{2}}), \label{speeq}
  \end{equation}
$r_{+}>r_{-}$, then the above specific heat is positive. The
stability condition is equivalent to the specific heat being
positive, so that the corresponding canonical ensemble is stable.
\section{Logarithmic correction to the Bekenstein-Hawking entropy and Cardy-Verlinde formula}
There has been much recent interest in calculating the quantum
corrections to $S_{BH}$ (the Bekenestein-Hawking entropy)
\cite{maj1x}-\cite{set5}. The corrected formula takes the form
\begin{equation}
\label{entro} {\cal S}=S_0-\frac{1}{2}{\rm ln }{C}+\ldots
\end{equation}
When $r_{+}\gg r_{-}$, $C\simeq S_{bh}= S_0$, in this case we have
\begin{equation}
\label{entro1} {\cal S}=S_0-\frac{1}{2}{\rm ln }{S_0}+\ldots
\end{equation}
\\
 It is now
possible to drive the corresponding correction to Cardy-Verlinde
formula. In a recent paper, Verlinde \cite{Verl} propound a
generalization of the Cardy formula which holds for the ($1+1$)
dimensional Conformal Field Theory (CFT), to $(n+1)$-dimensional
spacetime described by the metric \be
ds^{2}=-dt^{2}+R^{2}d\Omega_{n}\ee where $R$ is the radius of a
$n$-dimensional sphere.
\par\noindent The generalized
Cardy formula (hereafter named Cardy-Verlinde formula)  is given
by \be S_{CFT}=\frac{2\pi
R}{\sqrt{ab}}\sqrt{E_{C}\left(2E-E_{C}\right)} \label{cvf}\ee
where $E$ is the total energy, $E_{C}$ is the Casimir energy, $a$
and $b$ a priori arbitrary positive coefficients, independent of
$R$ and $S$. The definition of the Casimir energy is derived by
the violation of the Euler relation as \be E_{C}\equiv
n\left(E+pV-TS-J\Omega\right)\label{casimir1}\ee where the
pressure of the CFT is defined as $p=E/nV$. The total energy may
be written as the sum of two terms \be E(S, V)=E_{E}(S,
V)+\frac{1}{2}E_{C}(S, V)\label{ext}\ee where $E_{E}$ is the
purely extensive part of the total energy $E$. The Casimir energy
$E_{C}$ as well as the purely extensive part of energy $E_{E}$
expressed in terms of the radius $R$ and the entropy $S$ are
written as \bea
E_{C}&=&\frac{b}{2\pi R}S^{1-\frac{1}{n}}\label{cftcas1}\\
E_{E}&=&\frac{a}{4\pi
R}S^{1+\frac{1}{n}}\label{exten1}\hspace{1ex}.\eea After the work
of Witten on AdS$_{d}$/CFT$_{d-1}$ correspondence \cite{witten},
Savonije and Verlinde proved that the Cardy-Verlinde formula
(\ref{cvf}) can be derived using the thermodynamics of
AdS-Schwarzschild black holes in arbitrary dimension \cite{sav}.
 For the present discussion, the total entropy is assumed to be of
 the form Eq.(\ref{entro1}), where the uncorrected entropy, $S_{0}$
 correspondence to that associated in Eq. (\ref{entr1}).
Since the two-dimensional Ach\'ucarro-Ortiz black hole
 is asymptotically anti-de-Sitter, the total energy is $E=M$.
  It then follows by
 employing Eqs.(\ref{mass}-\ref{temp1}) that the Casimir energy Eq.(\ref{casimir1}) can
 be expressed in term of the uncorrected entropy.
 \begin{equation}
  E_{C} =\frac{J^{2}}{2r_{+}^{2}}-\frac{1}{2}T_{H} Ln S_{0},
\end{equation}
Then by setting the above corrected Casimir energy in Eq.(15) and
expanding  in term of $\frac{1}{2}T_{H} Ln S_{0}$ we obtain
\bea\label{sapro} \frac{2\pi
R}{\sqrt{ab}}\sqrt{E_{C}\left(2E-E_{C}\right)}\simeq
S_{0}(1+\frac{1}{2}T_{H}ln S_0 \frac{E_{C} -E}{E_{C}(2E-E_{C})}).
\eea In the limit where the correction is small, the coefficient
of the logarithmic term on the right-hand side of
Eq.(\ref{sapro}) can be expressed in terms of the energy and
Casimir energy
\begin{equation}
\frac{(E_{C}-E)}{2E_{C} (2E-E_{C})}T_H
S_0=\frac{(E_{C}-E)(2E-E_{C} -E_q)}{2E_{C}  (2E-E_{C}
)},\label{cof}
\end{equation}
where
\begin{equation}
E_q=Q\phi=J\Omega, \label{eeq}
\end{equation}
is the electromagnetic energy, in our analysis the charge $Q$ is
the angular momentum $J$ of the two-dimensional Achucarro-Oritz
black hole, the corresponding electric potential $\phi$ is the
angular velocity $\Omega$.
 We may
conclude, therefore that in the limit where the logarithmic
corrections are sub-dominant, Eq.(\ref{sapro}) can be rewritten to
express the entropy in terms of the energy,  and Casimir energy.
 \bea S_0=\frac{2\pi R}{\sqrt{ab}}\sqrt{E_{C}\left(2E-E_{C}\right)}
 -\frac{(E_C-E)(2E-E_C-E_q)}{2E_C(2E-E_C)}
 ln(\frac{2\pi R}{\sqrt{ab}}\sqrt{E_{C}\left(2E-E_{C}\right)})
 \eea and consequently, the total entropy Eq.(\ref{entro1}) to
first order in the logarithmic term, is given by  \bea
 S \simeq \frac{2\pi R}{\sqrt{ab}}\sqrt{E_{C}\left(2E-E_{C}\right)}
 -(\frac{(E_C-E)(2E-E_C-E_q)}{2E_C(2E-E_C)}+\frac{1}{2})
 ln(\frac{2\pi R}{\sqrt{ab}}\sqrt{E_{C}\left(2E-E_{C}\right)}) \eea
Therefore taking into account thermal fluctuations defines the
logarithmic corrections to the black hole entropies. As a result
the Cardy-Verlinde formula receive logarithmic corrections in our
interest  Achucarro-Oritz black hole background in two dimension,
in the way similar to the Cardy-Verlinde formula for the SAdS and
SdS black holes in 5-dimension \cite{{od},{od1}} also for TRNdS
black holes in any dimension\cite{set5}. It is easily seen that
the logarithmic prefactor is negative and therefore the thermal
corrections are also negative. Furthermore, the entropy of
two-dimensional Achucarro-Oritz black hole described in the
context of Das et al \cite{das2x} analysis by the modified
Cardy-Verlinde formula satisfy the holographic bound \cite{HOL}.

  \section{Conclusion}
 For a large class of black hole, the Bekenstein-Hawking entropy
 formula receives additive logarithmic corrections due to thermal
 fluctuations. On the basis of general thermodynamic arguments, Das et al
 \cite{das2x} deduced that the black hole entropy can be expressed as
\be {\cal S} = \ln \r = S_0 - \f{1}{2} \ln \le( C~T^2 \ri) +
\cdots \la{corr3}. \ee
 In this paper we have analyzed this correction of the entropy of Achucarro-Oritz
 black hole in two dimension in the light of AdS/CFT. We have obtain the logarithmic
 correction to  black hole entropy. Then using the form of the
 logarithmic correction Eq.(\ref{entro1}) we
 have derived the corresponding correction to the Cardy-Verlinde formula which relates the
 entropy of a certain CFT to its total energy and Casimir energy.
 The result of this paper is that the CFT entropy can be written
 in the form Eq.(25).

\end{document}